\documentclass[cernpreprint, atlasdraft=False, texlive=2020, USenglish, texmf]{atlasdoc}
 
\usepackage{atlaspackage}
\usepackage[backref=false]{atlasbiblatex}

\usepackage{atlascontribute}
 
\usepackage[unit=true, hepparticle=true, hepprocess=true, jetetmiss=true, misc=true]{atlasphysics}
 
\usepackage[output=true]{atlastodo}
 
\graphicspath{{logos/}{figures/}}
 
\usepackage{ANA-STDM-2019-09-PAPER-defs}
\usepackage{multirow}
\usepackage{siunitx}

\AtlasTitle{Observation of $WWW\xspace$\ Production in $pp$ Collisions at $\sqrt{s} = \SI{13}{TeV}$ with the ATLAS Detector}

\PreprintIdNumber{CERN-EP-2021-243}
 
\AtlasAbstract{
This Letter reports the observation of $WWW\xspace$\ production and a measurement of its cross section using \SI{139}{$\text{fb}^{-1}$}\ of proton-proton collision data recorded at a center-of-mass energy of \SI{13}{TeV} by the ATLAS detector at the Large Hadron Collider. Events with two same-sign leptons (electrons or muons) and at least two jets, as well as events with three charged leptons, 
are selected. A multivariate technique is then used to discriminate between signal and background events. Events from $WWW\xspace$\ production are observed with a significance of 8.0 standard deviations, where the expectation is 5.4 standard deviations. The inclusive $WWW\xspace$\ production cross section is measured to be $820 \pm 100\,\text{(stat)} \pm 80\,\text{(syst)}$\,fb, approximately 2.6 standard deviations from the predicted cross section of $511 \pm 18$\,fb calculated at next-to-leading-order QCD and leading-order electroweak accuracy.
}
 
\author{The ATLAS Collaboration}

\AtlasJournal{\PRL}
 
\AtlasJournalRef{Phys. Rev. Lett. 129 (2022) 061803}
\AtlasDOI{10.1103/PhysRevLett.129.061803}

\hypersetup{pdftitle={ATLAS document},pdfauthor={The ATLAS Collaboration}}
 
\begin{document}
 
\maketitle

Measurements of triboson production at colliders directly probe the strength of gauge boson self-interactions within the standard model (SM) via triple gauge couplings and quartic gauge couplings
~\cite{BELANGER1992201, Godfrey:1995}.
Any significant deviations from the SM predictions would provide evidence of new physics at a higher energy scale than is presently accessible~\cite{Ren:2014bza, Kuang:2015gka, Agashe:2016rle, Agashe:2016kfr, Agashe:2017wss, Aguilar-Saavedra:2017iso}.
Triboson final states are among the least-understood SM processes due to their small production cross sections. In particular, searches for the production of three \PW bosons
(\www) have been performed by both the ATLAS~\cite{STDM-2015-07, STDM-2017-22} and CMS~\cite{CMS-SMP-17-013, CMS-SMP-19-014} Collaborations.
Using proton-proton ($pp$) collisions at a center-of-mass energy ($\sqrt{s}$) of \SI{13}{\TeV} delivered by the Large Hadron Collider (LHC)~\cite{Evans:2008zzb}, the ATLAS Collaboration analyzed \SI{80}{\ifb} of data and provided evidence for both \www\ and $WWZ$/$WZZ$ production~\cite{STDM-2017-22}, and the CMS Collaboration analyzed \SI{137}{\ifb} of data and observed the combined production of three massive vector bosons (\www, $WWZ$, $WZZ$ and $ZZZ$)~\cite{CMS-SMP-19-014}.
 
\begin{figure}[h]
\begin{center}
\includegraphics[width=0.95\textwidth]{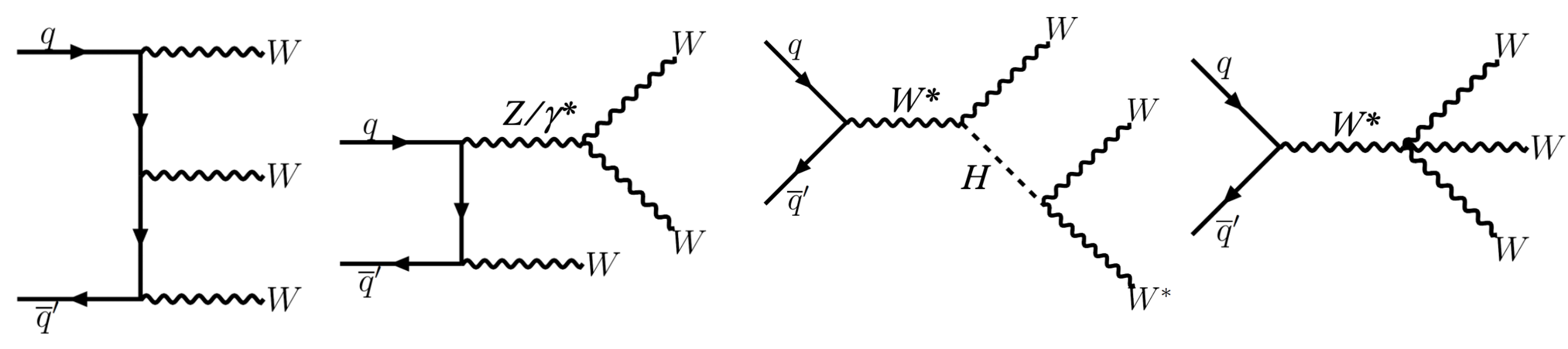}
\end{center}
\caption{Representative Feynman diagrams at LO for the production of three massive $W\xspace$\ bosons, including diagrams sensitive to triple and quartic gauge couplings.}
\label{fig:diagrams}
\end{figure}

This Letter reports the observation of \www\ production and a measurement of its cross section using \SI{139}{\ifb} of data at $\sqrt{s} = \SI{13}{\TeV}$~\cite{ATLAS-CONF-2019-021}
taken with the ATLAS detector.
At leading order (LO) in QCD, \www\ production can proceed via the radiation of each $W$ boson from a fermion, via a $W$ boson produced in association with an intermediate $\PZ/\gamma^*$ or
Higgs boson that decays via the $WW^*$ intermediate state,
or via a quartic gauge coupling vertex.
Representative Feynman diagrams are shown in Figure~\ref{fig:diagrams}.
The analysis selection is sensitive to processes with both on-shell and off-shell $W$ bosons decays. For simplicity all these processes (including $WH \rightarrow WWW^*$) are generically referred to as $WWW$ throughout this Letter.
Two decay channels, \twoltwojch and \threelch with $\ell= e$ or $\mu$, are considered and are hereafter referred to as 2$\ell$ and 3$\ell$, respectively.
Events with electrons and muons produced through $\tau$ leptons are also included.
The experimental signature of the 2$\ell$ channel consists of two same-sign charged leptons, missing transverse momentum, and two jets,
while the signature of the 3$\ell$ channel consists of three charged leptons
and missing transverse momentum.
 
The ATLAS detector~\cite{PERF-2007-01} is a multipurpose particle physics detector with cylindrical geometry.\footnote{ATLAS uses a right-handed coordinate system with its origin at the nominal interaction point (IP) in the center of the detector and the \(z\)-axis along the beam pipe.
The \(x\)-axis points from the IP to the center of the LHC ring, and the \(y\)-axis points upwards.
Cylindrical coordinates \((r,\phi)\) are used in the transverse plane, \(\phi\) being the azimuthal angle around the \(z\)-axis.
The pseudorapidity is defined in terms of the polar angle \(\theta\) as \(\eta = -\ln \tan(\theta/2)\).
Momentum in the transverse plane is denoted by \pT.}
It consists of an inner tracker (ID) surrounded by a superconducting solenoid, sampling electromagnetic (EM) and hadronic calorimeters, and a muon spectrometer (MS) with three toroidal superconducting magnets.
A two-level trigger system is used to select events for storage.
Events used in this analysis were selected online by single-electron or single-muon triggers~\cite{TRIG-2016-01, TRIG-2018-05, ATL-DAQ-PUB-2019-001}. An extensive software suite~\cite{ATL-SOFT-PUB-2021-001} is used in the reconstruction and analysis of real and simulated data, in detector operations, and in the trigger and data acquisition systems of the experiment.
 
The proton interaction vertex with the highest $p_T^2$ sum of associated ID tracks is selected as the primary vertex. Electrons are reconstructed from energy deposits in the EM calorimeter associated with tracks found in the ID.
Muons are reconstructed by combining tracks reconstructed in the ID with tracks or track segments found in the MS.
Electrons (muons) must have $\pT > \SI{20}{\GeV}$ and be reconstructed within $|\eta|<2.47$ ($|\eta|<2.5$), excluding electrons within $1.37<|\eta|<1.52$.
To ensure that selected leptons originate from the primary vertex, their tracks are required to have $|d_0/\sigma_{d_0}| < 5~(3)$ for electrons (muons) and $|z_0 \sin \theta|< \SI{0.5}{\mm}$ for both lepton flavors, where $d_0$ and $\sigma_{d_0}$ are the transverse impact parameter and its uncertainty, and $z_0$ is the longitudinal impact parameter.
Electrons are required to satisfy the ``tight'' likelihood-based identification criterion defined in Ref.~\cite{EGAM-2018-01}, and muons must satisfy the ``medium'' cut-based identification criterion defined in Ref.~\cite{MUON-2018-03}.
To reject leptons that likely originate from light-hadron decays or heavy-flavor decays, leptons are required to pass a tight isolation requirement (``PLVTight'')~\cite{HIGG-2017-02},
which takes into account the energy deposits and charged-particle tracks (including the lepton track) in a cone around the lepton direction.
Electrons must also satisfy a charge identification criterion based on a boosted decision tree (BDT) discriminant~\cite{PERF-2017-01} to reduce the contamination from electrons with misidentified electric charge.
 
Particle-flow jets are reconstructed from tracks in the ID and topological energy clusters in the \mbox{calorimeter~\cite{PERF-2015-09}.}
Jet candidates are required to have $\pT > \SI{30}{\GeV}$ in the forward region ($2.5<|\eta|<4.5$) and $\pT > \SI{20}{\GeV}$ in the central region ($|\eta|<2.5$).
To reduce the effect from additional $pp$ collisions in the same or
a nearby bunch crossing (pileup), jets with $\SI{20}{\GeV} < \pT < \SI{60}{\GeV}$ and $|\eta| < 2.5$
are required to pass a ``jet vertex tagger'' requirement~\cite{PERF-2014-03}.
Jets containing \Pqb-flavored hadrons (``\Pqb jets'') are identified by a multivariate
discriminant~\cite{FTAG-2018-01, ATLAS-CONF-2018-006} combining track impact parameter values
with information from secondary vertices reconstructed within the jet.
A working point corresponding to an 85\% efficiency for identifying \Pqb jets in \ttbar events is used.
Procedures described in Ref.~\cite{STDM-2017-22} that ensure the selected electron, muon, and jet candidates
do not overlap are applied before the lepton ``PLVTight'' and $|d_0/\sigma_{d_0}|$ requirements.
 
The missing transverse momentum, whose magnitude is denoted by \met, is calculated as the negative of the vector sum of the transverse momenta of all reconstructed objects associated with the primary vertex.
To account for soft hadronic activity, a term including tracks associated with the primary vertex but not with any of the reconstructed objects is included in the calculation of \met~\cite{PERF-2016-07}.
 
To select candidates in the 2$\ell$ signal regions (SRs), events are required to have exactly two leptons with the same electric charge, at least two central jets, and no identified \Pqb jets.
Three final states based on the lepton flavors are considered, namely \ee, \me, and \mm.
The highest-\pT lepton must have $\pT>\SI{27}{\GeV}$ and the dilepton invariant mass
$m_{\ell\ell}$ is required to be between \SI{40}{\GeV} and \SI{400}{\GeV}. The two highest-\pT
central jets are required to
have $m_{jj}< \SI{160}{\GeV}$ and $|\Delta\eta_{jj}|< 1.5$, where $m_{jj}$ is the dijet invariant mass
and $\Delta\eta_{jj}$ is the pseudorapidity separation between the two jets.
The $m_{jj}$ and $\Delta\eta_{jj}$ selection suppresses contributions from the $\Wpm\Wpm$ vector-boson scattering process.
In the case of the \ee final state, the dilepton system is required to have $m_{\ell\ell}< \SI{80}{\GeV}$ or $m_{\ell\ell}> \SI{100}{\GeV}$ and an \met significance~\cite{ATLAS-CONF-2018-038} requirement, $\mathcal{S}(\met)>3$, is applied to suppress contributions from the $Z+$jets process where the charge of one electron is misidentified.
To select candidates in the 3$\ell$ SR, events are required to have exactly three leptons
including at least one with $\pT > \SI{27}{\GeV}$, no identified \Pqb jets, and no same-flavor opposite-sign (SFOS) lepton pairs. Events with \eem and \mme final states are considered.
To suppress contributions from $WZ+$jets production in the 2$\ell$ SRs and $ZZ+$jets production in the 3$\ell$ SR, events are removed if they contain additional electrons (muons)
reconstructed with $\pT > 7~(4.5)\,\si{\GeV}$ and $|\eta|<2.47~(2.7)$ passing the loose~\cite{EGAM-2018-01}~(\cite{MUON-2018-03}) identification requirement.
 
Monte Carlo (MC) simulated samples are used to model the signal \www\ process, as well as contributions from other physics processes with prompt leptons.
Simulated events were processed through the full ATLAS detector simulation~\cite{SOFT-2010-01} based on $\GEANT$~\cite{Agostinelli:2002hh}.
The effects of pileup are included in the simulation.
 
Events with three on-shell \PW bosons were generated
by \SHERPA[2.2.2]~\cite{Gleisberg:2008ta, Hoeche:2009rj} with
the \NNPDF[3.0nnlo] parton distribution function (PDF)~\cite{Ball:2014uwa}.
Events with an off-shell $W$ boson through $WH \rightarrow WWW^*$ were generated using \POWHEGBOX[v2]~\cite{Luisoni:2013cuh} interfaced to \PYTHIA[8.235]~\cite{Sjostrand:2007gs}
for parton showering~\cite{Mrenna:2016sih} with the \NNPDF[2.3lo] PDF and the AZNLO set of
tuned parameters~\cite{STDM-2012-23}.
Both processes are included in the signal definition and were generated at next-to-leading-order (\NLO) QCD accuracy and LO electroweak accuracy with all spin correlations taken into account in the vector-boson decays.
The cross section for the process with on-shell (off-shell $WH\rightarrow WWW^*$) $W$ bosons is $209 \pm 17$ fb ($302 \pm 8$ fb). The inclusive cross section is $\sigma^{\text{pred.}}(pp \rightarrow WWW) = 511 \pm 18$ fb.
The cross sections and uncertainties used in this analysis are consistent with the latest calculations with NLO QCD and NLO electroweak corrections applied for the on-shell $WWW$ process~\cite{Dittmaier:2017bnh} and with next-to-next-to-next-to-leading-order (NNNLO) QCD and NLO electroweak corrections applied for the $WH\rightarrow WWW^*$ process~\cite{Anastasiou:2016cez}.
 
The dominant background originates from the $\ell\nu\ell\ell+\text{jets}(WZ+\text{jets})$ process, and its contribution is estimated using simulated events generated with \SHERPA[2.2.2] using the \NNPDF[3.0nnlo] PDF and a threshold of \SI{4}{GeV} on the $Z$ boson mass.
The matrix element calculations for the $WZ$ process were performed with up to one additional parton at \NLO QCD accuracy and up to three additional partons at \LO QCD accuracy.
To ensure proper modeling of the $WZ$ background, the MC predictions for $WZ+0$~jets, $WZ+1$~jet, and $WZ~+\ge 2~\text{jets}$ are multiplied by scale factors obtained during the fit to the data to be described later, which includes three $WZ$ control regions (CR).
These CRs are obtained by requiring exactly three leptons with one SFOS lepton pair, \met significance $\mathcal{S}(\met)>3$, no \Pqb jets identified, and a trilepton invariant mass $\SI{110}{\GeV} <m_{\ell\ell\ell}<  \SI{500}{\GeV}$.
 
The contribution from backgrounds with nonprompt leptons from hadron (including $b$-flavored and $c$-flavored hadrons) decays and jets misidentified as leptons
is estimated using a data-driven method described in Ref.~\cite{STDM-2014-05}.
Lepton-like jets are defined by requiring the leptons to meet a looser selection criterion but fail the signal-lepton requirements.
Compared to signal leptons, muonlike jets have $|d_0/\sigma_{d_0}| < 10$, and electronlike jets have the likelihood-based identification criterion loosened to ``medium''~\cite{EGAM-2018-01}.
The PLVTight tight isolation criterion is dropped for both lepton flavors.
Since the nonprompt-lepton background in the SRs comes mainly from
the \ttbar process where one of the \Pqb jets is misidentified as an isolated lepton, a lepton ``fake factor'' is determined from \ttbar-enriched samples selected using the same
signal region criteria, except requiring one \Pqb jet and, in the 3$\ell$ case, including events with a SFOS lepton pair with $m_{\ell\ell}<\SI{80}{\GeV}$ or $m_{\ell\ell}>\SI{100}{\GeV}$.
To estimate the nonprompt-lepton background, this fake factor is applied as a weight to
events selected with the same set of criteria as the signal region
but with a lepton like jet.
The nonprompt-lepton background estimate is validated by checking that the estimate agrees with the data in the $t\bar{t}$-enriched samples where the fake factors are measured.
 
The $W\gamma/Z\gamma$ background mostly contributes to the event selection when the photon is being misidentified as an electron.
This contribution (referred to as ``$\gamma$ conversions'') is evaluated using a data-driven method similar to the nonprompt-lepton background estimation by introducing electronlike photons.
An electronlike photon is a reconstructed object that is like an electron except that its associated track has no hits
in the innermost layer of the pixel detector. The photon fake factor is determined using
$Z\gamma \rightarrow \mu\mu\gamma$ events selected
with two muons, no \Pqb jets, and one electron or one electronlike photon. The trilepton
invariant mass must satisfy $\SI{80}{\GeV} < m_{e\mu\mu} < \SI{100}{\GeV}$.
 
The charge-flip background originates from processes in which the charge of at least one prompt electron is misidentified.
The muon charge misidentification rate is found to be negligible.
The electron charge misidentification rate is measured using a tag-and-probe method applied to $Z \rightarrow e^+ e^-$ events, where the two electrons have the same reconstructed charge~\cite{STDM-2014-05}.
The charge-flip background is estimated by applying the measured electron charge misidentification rates to \ee, $e^\pm \mu^\pm$, and $e^\pm e^\pm \mu^\mp$ data events that meet all signal region requirements except for the SFOS lepton pair veto requirement.
This method is validated with events selected using the same set of signal region criteria as used in the \ee final state,
except the dilepton mass must satisfy $\SI{80}{\GeV} < m_{\ell\ell} < \SI{100}{\GeV}$ and the \met\ significance requirement $\mathcal{S}(\met)>3$ is removed.
 
Other SM processes with prompt leptons include $t\bar{t} W$, $t\bar{t} Z$, $tZq$, $t\bar{t}H$, $WWZ$, $WZZ$, $ZZZ$, and $W^\pm W^\pm jj$ production. Their contributions
are estimated using simulated events normalized to the integrated luminosity of the data sample and the cross sections provided by the event generators.
The $t\bar{t} W$, $t\bar{t} Z$, and $tZq$ processes were modeled using \MGNLO[2.3.3]~\cite{Alwall:2014hca} together with \PYTHIA[8.210], with up to two additional partons in the matrix-element calculations.
The $t\bar{t}H$ process was modeled using \POWHEGBOX[v2] interfaced with \PYTHIA[8.230].
Other triboson processes ($WWZ$, $WZZ$, $ZZZ$) and the strong production of $W^\pm W^\pm jj$ were modeled using \SHERPA[2.2.2].
The calculations for triboson processes were performed with no extra partons at NLO QCD accuracy and up to two additional partons at LO QCD accuracy. The electroweak production of $W^\pm W^\pm jj$ was modeled using \SHERPA[2.2.11], and the calculations were performed with up to one additional parton at LO QCD accuracy.
Contributions from the on-shell $WWW$ and $WH\rightarrow WWW^*$ processes were removed from $W^\pm W^\pm jj$ production. Contributions from double parton scattering processes are found to be negligible.
 
To improve the separation between signal and background, two BDTs are trained using the XGBoost~\cite{Chen:2016:XST:2939672.2939785} package and are applied separately to the 2$\ell$ and 3$\ell$ SRs.
All backgrounds are included in the BDT training.
Each BDT is trained with 11 variables, some of which differ between the two sets.
The three variables with the highest discriminating power are:
$|m_{jj}-m_W|$, where $m_W$ is the pole mass of the $W$ boson, forward jet $p_T$, and $\mathcal{S}(\met)$ for the $2\ell$ channnel, and the ratio $\mathcal{S}(\met)/\met$, $p_T$ of the second highest-$p_T$ lepton, and number of jets for the $3\ell$ channel.
A $k$-fold cross-validation procedure is used to produce the final discriminant.
Fivefold (fourfold) cross-validation BDTs are trained in the 2$\ell$ SRs (3$\ell$ SR) and each BDT is trained on 80\% (75\%) of the expected signal and background events. To produce the BDT distribution used in the fit, each of the five (four) trained BDTs is applied to the 20\% (25\%) of the events not used to train the BDTs.
 
To extract the \www\ inclusive cross section, a binned maximum likelihood fit~\cite{Cranmer:2012sba} is performed using the BDT distributions in the 2$\ell$ SRs (\ee, \me, and \mm) and the 3$\ell$ SR as well as the $m_{\ell\ell\ell}$ distributions in the three $WZ$ CRs, amounting to seven distributions with \num{50} bins in total.
The fit includes four unconstrained parameters that scale the number of events for a particular process predicted by MC simulation: the signal strength $\mu(WWW)$ for $WWW$ production and three scale factors for $WZ+0$~jets, $WZ+1$~jet and $WZ~+\ge 2$~jets.
The ratio of on-shell \www\ production to $WH \rightarrow WWW^*$ production is determined from MC simulation and is allowed to vary within the theoretical uncertainties of the two processes.
 
\begin{figure}[h!]  
\begin{center}
\includegraphics[width=0.35\textwidth]{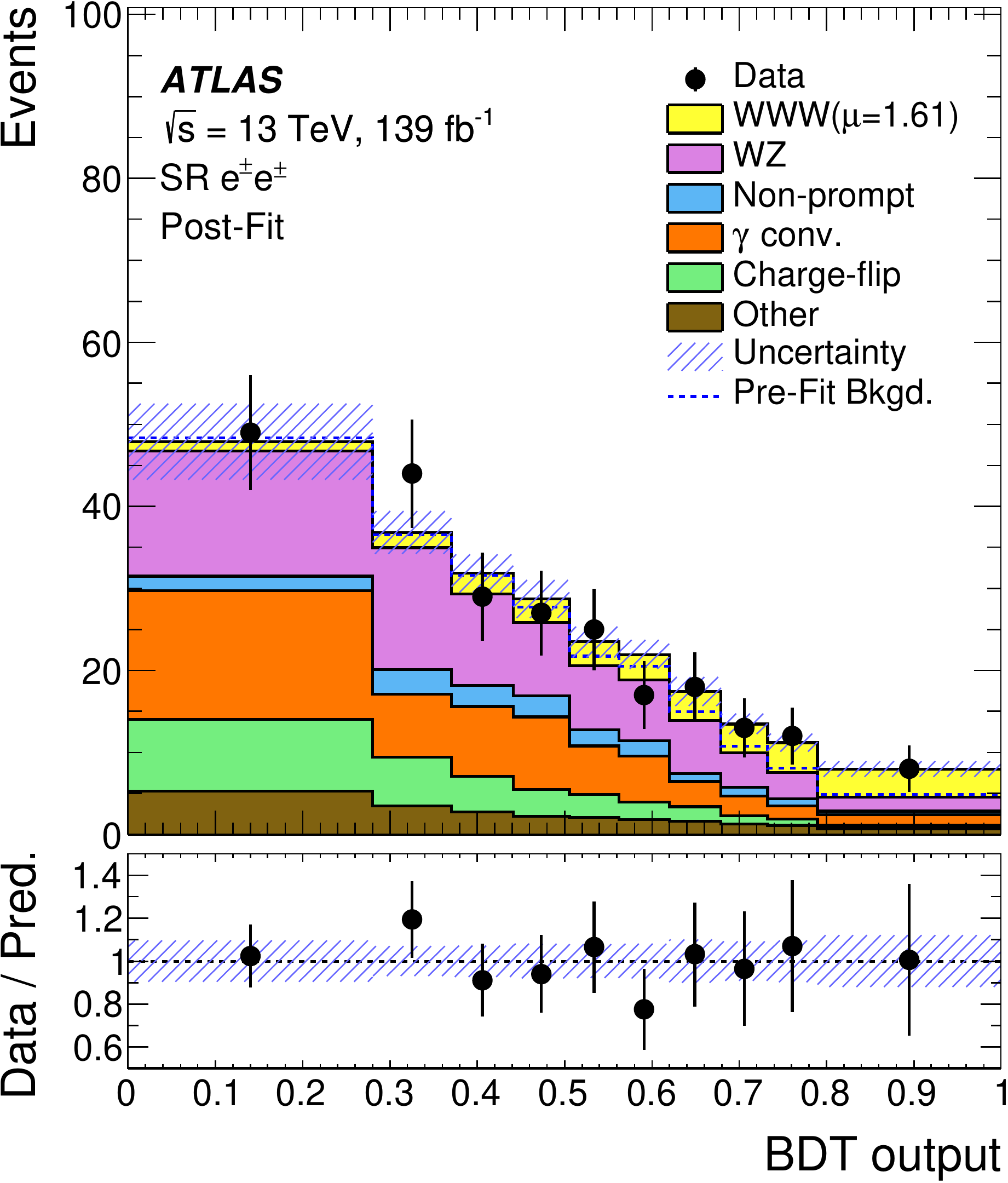}
\includegraphics[width=0.35\textwidth]{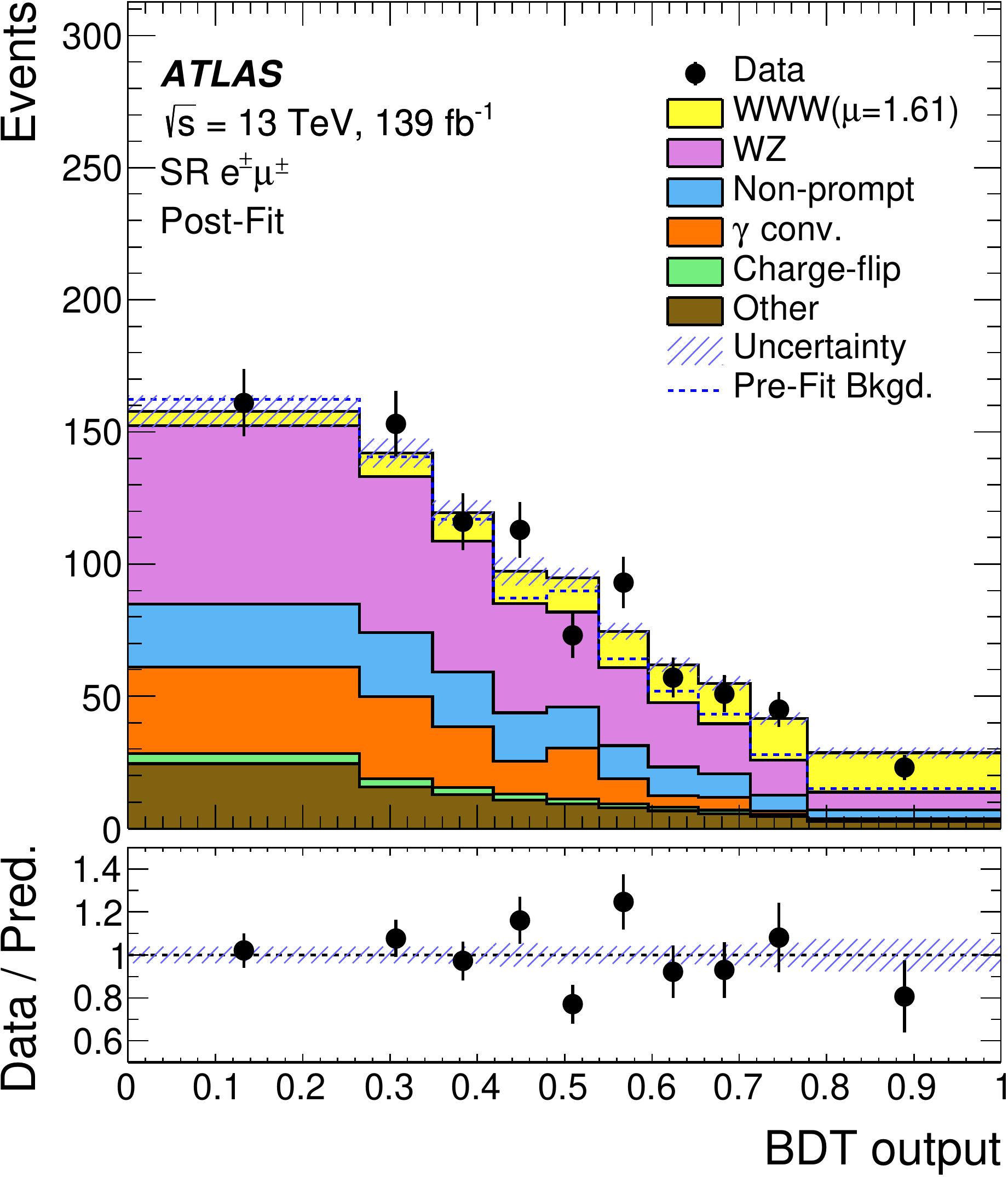}\\
\includegraphics[width=0.35\textwidth]{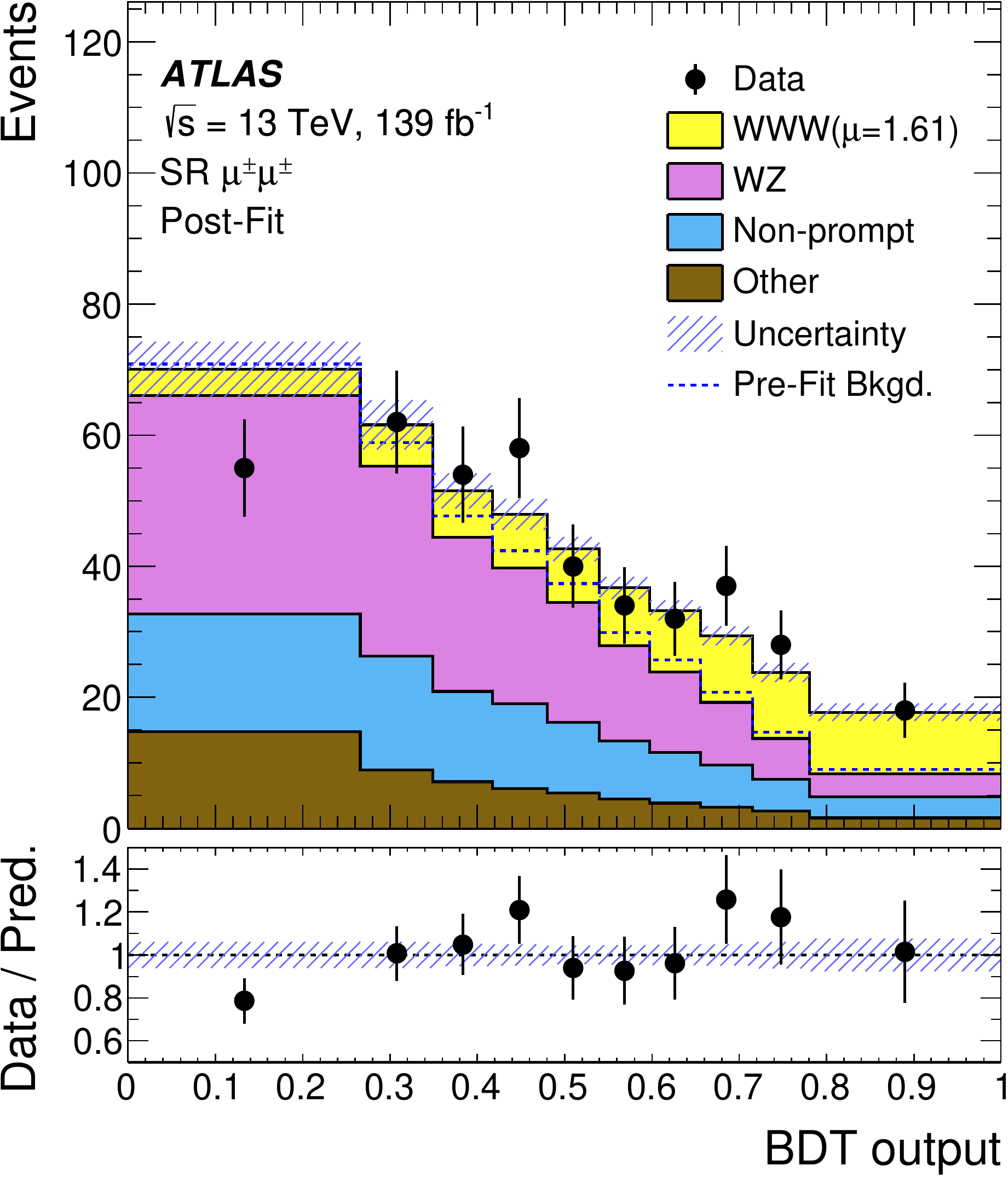}
\includegraphics[width=0.35\textwidth]{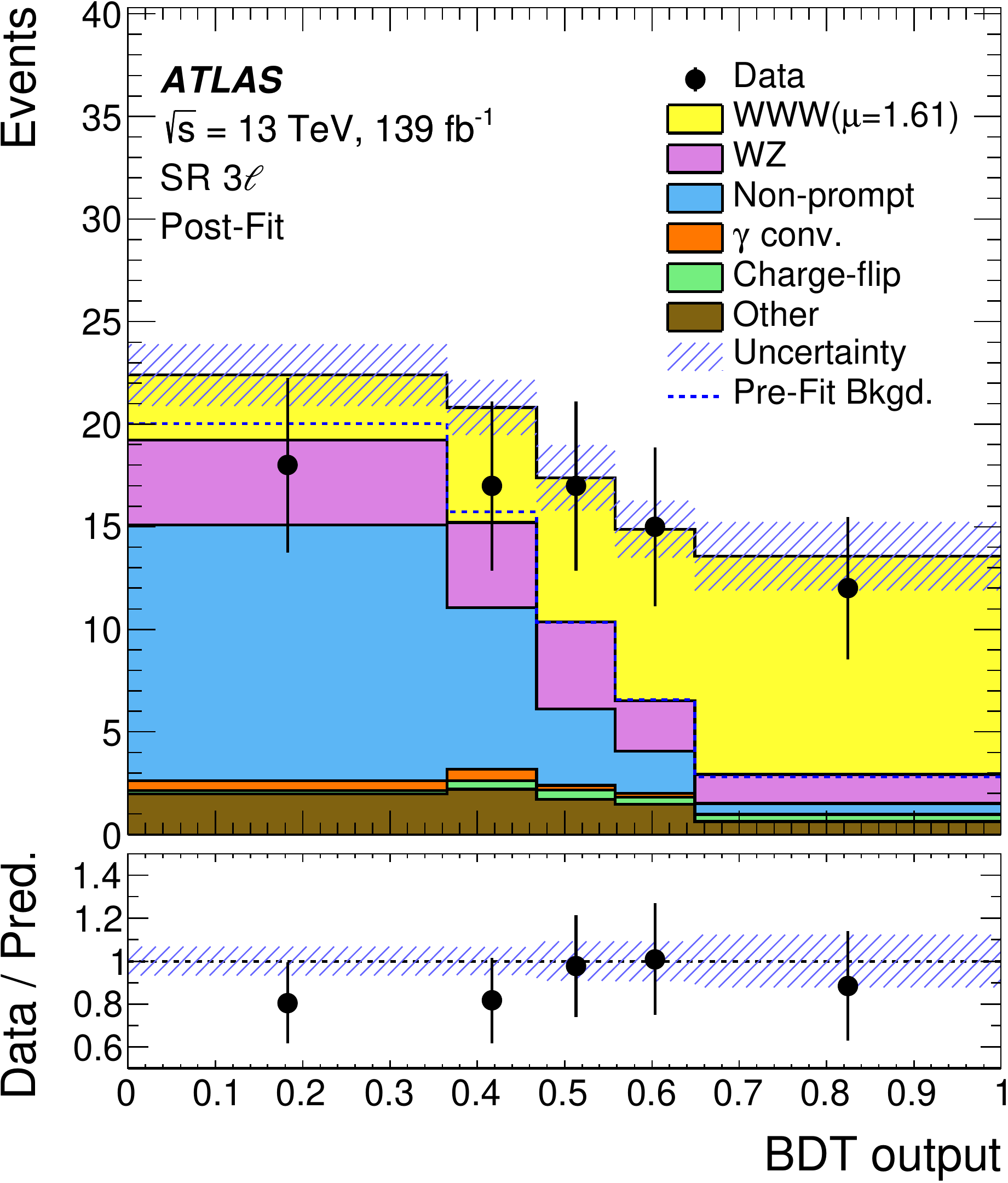}
\end{center}
\caption{Postfit BDT output distribution in the $e^{\pm}e^{\pm}\xspace$\ (top left), $e^{\pm}\mu^{\pm}\xspace$\ (top right), $\mu^{\pm}\mu^{\pm}\xspace$\ (bottom left) and $3\ell\xspace$ (bottom right) channels.
The bottom panel of each plot shows the ratio of the data to the total prediction.
The uncertainty bands include both the statistical and systematic uncertainties as obtained by the fit. The signal is scaled to the fitted signal strength of 1.61.}
\label{fig:results}
\end{figure}

Systematic uncertainties are included in the fit as nuisance parameters constrained by Gaussian probability density functions. Correlations between systematic uncertainties arising from common sources are maintained across processes and channels.
Instrumental systematic uncertainties are related to the lepton trigger,
reconstruction and identification efficiencies~\cite{MUON-2018-03, PERF-2017-01}, lepton isolation criteria~\cite{HIGG-2017-02}, lepton energy
scale and resolution~\cite{MUON-2018-03, PERF-2017-03}, jet energy scale and resolution~\cite{JETM-2018-05}, jet vertex tagging~\cite{ATLAS-CONF-2014-018, PERF-2016-06}, \Pqb jet identification~\cite{FTAG-2018-01}, modeling of \met~\cite{Aaboud:2018tkc} and pileup,
and integrated luminosity~\cite{ATLAS-CONF-2019-021,LUCID2}.
Theoretical uncertainties associated with the signal processes and the background processes with prompt leptons are evaluated using simulation.
For the signal and $WZ$ background, acceptance and distribution shape uncertainties due to the
renormalization and factorization scales~\cite{ATL-PHYS-PUB-2016-002}, PDFs~\cite{Butterworth:2015oua},
and parton showering, are also included in the simultaneous fit.
The normalization uncertainties for the processes included in the ``Other'' background category in Table~\ref{tab:yields} and Figure~\ref{fig:results} are between 10\% and 20\%~\cite{LHCHiggsCrossSectionWorkingGroup:2016ypw, TOPQ-2018-01, STDM-2017-06, TOPQ-2016-11}.
The fit includes the systematic uncertainties of each of the data-driven background estimates, and also the systematic uncertainties due to limited MC sample size.
 
The signal strength is measured to be $\mu(WWW) = 1.61 \pm 0.25$, where the uncertainty also includes the signal cross-section uncertainty (3.6\%) affecting the predicted inclusive cross section from the signal MC samples.
The three $WZ$ scale factors are found to be $1.12 \pm 0.11$, $0.98 \pm 0.04$ and $0.88 \pm 0.18$ for the 0-jet, 1-jet, and ${\geq}2$-jet bins.
Table~\ref{tab:yields} shows the postfit signal and background event yields as well as the observed yield in each SR.
The contribution of the $WH$ process to the \www\ yield ranges between 40\% and 44\% in the four SRs.
All nuisance parameters remain within their one standard deviation uncertainty after the fit.
Figure~\ref{fig:results} shows a comparison between data and postfit predictions for the BDT output score distribution in all SRs.
For various postfit kinematic distributions in the SRs, data and predictions are found to have a $p$ value greater than 0.05 from a $\chi^2$ test that takes into account the systematic uncertainties and correlations used in the fit to data.

\begin{table}[h!] 
\begin{center}
\caption{
Number of events for postfit signal, background and data observed in the 2$\ell$ and 3$\ell$ SRs.
The uncertainties shown include both the statistical and systematic contributions.
}
\sisetup{round-mode=places, round-precision=1}
\begin{tabular}{l|
*4{l@{$\,\pm\,$}l}
}
\toprule
& \multicolumn{2}{c}{$e^{\pm}e^{\pm}$}
& \multicolumn{2}{c}{$e^{\pm}\mu^{\pm}$}
& \multicolumn{2}{c}{$\mu^{\pm}\mu^{\pm}$}
& \multicolumn{2}{c}{$3\ell$}\\
\midrule
\www\ signal
& \numRFT{3.1}{3}{28.419}  & \numRFT{2.1}{2}{4.346}
& \numRFT{3.0}{3}{124.164} & \numRFT{2.0}{2}{18.745}
& \numRFT{3.0}{2}{81.597}  & \numRFT{2.0}{2}{12.400}
& \numRFT{3.1}{3}{34.793}  & \numRFT{1.1}{2}{5.191}
\\
$WZ$
& \numRFT{3.1}{3}{81.103} & \numRFT{2.1}{2}{5.656}
& \numRFT{3.0}{3}{345.687} & \numRFT{2.0}{2}{22.104}
& \numRFT{3.0}{2}{170.952} & \numRFT{2.0}{1}{9.759}
& \numRFT{3.1}{3}{16.392} & \numRFT{1.1}{2}{1.453}
\\
Charge-flip
& \numRFT{3.1}{3}{31.064} & \numRFT{2.1}{2}{7.302}
& \numRFT{3.0}{2}{19.225} & \numRFT{2.0}{1}{4.568}
& \multicolumn{2}{c}{-}
& \numRFT{3.1}{2}{1.719}  & \numRFT{1.1}{1}{0.409}
\\
$\gamma$ conversions
& \numRFT{3.1}{3}{60.846}  & \numRFT{2.1}{2}{8.536}
& \numRFT{3.0}{3}{139.022} & \numRFT{2.0}{2}{14.942}
& \multicolumn{2}{c}{-}
& \numRFT{3.1}{2}{1.511}   & \numRFT{1.1}{1}{0.120}
\\
Nonprompt
& \numRFT{3.1}{3}{17.038}  & \numRFT{2.1}{2}{4.023}
& \numRFT{3.0}{3}{144.615} & \numRFT{2.0}{2}{23.284}
& \numRFT{3.0}{3}{103.975} & \numRFT{2.0}{2}{20.961}
& \numRFT{3.1}{3}{26.635}  & \numRFT{1.1}{2}{2.882}
\\
Other
& \numRFT{3.1}{3}{22.315} & \numRFT{2.1}{2}{2.378}
& \numRFT{3.0}{2}{99.944} & \numRFT{2.0}{1}{9.911}
& \numRFT{3.0}{2}{58.001} & \numRFT{2.0}{1}{5.987}
& \numRFT{3.1}{2}{7.965}  & \numRFT{1.1}{1}{0.882}
\\
\midrule
Total predicted
& \numRFT{3.0}{3}{240.785} & \numRFT{2.1}{2}{10.608}
& \numRFT{3.0}{3}{872.657} & \numRFT{2.0}{2}{22.025}
& \numRFT{3.0}{3}{414.526} & \numRFT{2.0}{2}{17.146}
& \numRFT{3.1}{3}{89.014}  & \numRFT{1.1}{2}{5.444}
\\
\midrule
Data
& \multicolumn{2}{c}{242}
& \multicolumn{2}{c}{885}
& \multicolumn{2}{c}{418}
& \multicolumn{2}{c}{79}
\\
\bottomrule
\end{tabular}
\label{tab:yields}
\end{center}
\end{table}

The background-only hypothesis is rejected with an observed (expected) significance of 6.6 (4.0) standard deviations
for the 2$\ell$ SRs and 4.8 (3.8) standard deviations for the 3$\ell$ SR calculated using the asymptotic approximation~\cite{Cowan:2010js}. The combined observed (expected) significance is found to be \obssigma (\expsigma) standard deviations, constituting the first observation of \www\ production.
The signal strength is also measured separately by fitting the BDT distribution in each SR with the three $WZ$ CRs.
The values are found to be consistent: $1.54 \pm 0.76$ for \ee, $1.44 \pm 0.39$ for \me, $2.23 \pm 0.46$ for \mm, and $1.32 \pm 0.39$ for 3$\ell$.
 
The measured inclusive $pp \rightarrow \www$ production cross section is calculated as the product of the
measured signal strength and the cross section from MC simulation, and is found to be
$\sigma^{\text{meas.}}(pp \rightarrow WWW)=820 \pm 100~{\text{(stat)}} \pm 80~{\text{(syst)}}$~fb.
The largest systematic uncertainty contribution is 6\% from data-driven estimates (mainly nonprompt background), followed by 3\% from prompt-lepton-background modeling uncertainties (primarily $WZ$ theory uncertainties).

\FloatBarrier
 
In conclusion, the first observation of the $pp \rightarrow \www$ process is reported by the ATLAS experiment at the LHC. Events with two same-sign charged leptons in association with at least two jets, as well as events with three charged leptons and no same-flavor opposite-sign lepton pairs, were selected from 139~fb$^{-1}$ of 13~\TeV\ $pp$ collisions. Two BDTs were trained to improve the separation between signal and background. The SM background-only hypothesis is rejected with an observed (expected) significance of \obssigma (\expsigma) standard deviations.
The inclusive $pp \rightarrow \www$ production cross section is measured to be $820 \pm 100~{\text{(stat)}} \pm 80~{\text{(syst)}}$~fb,  approximately 2.6 standard deviations from the predicted cross section of $511 \pm 18$~fb calculated at NLO QCD and LO electroweak accuracy.
\section*{Acknowledgments}

We thank CERN for the very successful operation of the LHC, as well as the
support staff from our institutions without whom ATLAS could not be
operated efficiently.
 
We acknowledge the support of
ANPCyT, Argentina;
YerPhI, Armenia;
ARC, Australia;
BMWFW and FWF, Austria;
ANAS, Azerbaijan;
SSTC, Belarus;
CNPq and FAPESP, Brazil;
NSERC, NRC and CFI, Canada;
CERN;
ANID, Chile;
CAS, MOST and NSFC, China;
Minciencias, Colombia;
MEYS CR, Czech Republic;
DNRF and DNSRC, Denmark;
IN2P3-CNRS and CEA-DRF/IRFU, France;
SRNSFG, Georgia;
BMBF, HGF and MPG, Germany;
GSRI, Greece;
RGC and Hong Kong SAR, China;
ISF and Benoziyo Center, Israel;
INFN, Italy;
MEXT and JSPS, Japan;
CNRST, Morocco;
NWO, Netherlands;
RCN, Norway;
MEiN, Poland;
FCT, Portugal;
MNE/IFA, Romania;
JINR;
MES of Russia and NRC KI, Russian Federation;
MESTD, Serbia;
MSSR, Slovakia;
ARRS and MIZ\v{S}, Slovenia;
DSI/NRF, South Africa;
MICINN, Spain;
SRC and Wallenberg Foundation, Sweden;
SERI, SNSF and Cantons of Bern and Geneva, Switzerland;
MOST, Taiwan;
TAEK, Turkey;
STFC, United Kingdom;
DOE and NSF, United States of America.
In addition, individual groups and members have received support from
BCKDF, CANARIE, Compute Canada and CRC, Canada;
COST, ERC, ERDF, Horizon 2020 and Marie Sk{\l}odowska-Curie Actions, European Union;
Investissements d'Avenir Labex, Investissements d'Avenir Idex and ANR, France;
DFG and AvH Foundation, Germany;
Herakleitos, Thales and Aristeia programmes co-financed by EU-ESF and the Greek NSRF, Greece;
BSF-NSF and GIF, Israel;
Norwegian Financial Mechanism 2014-2021, Norway;
NCN and NAWA, Poland;
La Caixa Banking Foundation, CERCA Programme Generalitat de Catalunya and PROMETEO and GenT Programmes Generalitat Valenciana, Spain;
G\"{o}ran Gustafssons Stiftelse, Sweden;
The Royal Society and Leverhulme Trust, United Kingdom.
 
The crucial computing support from all WLCG partners is acknowledged gratefully, in particular from CERN, the ATLAS Tier-1 facilities at TRIUMF (Canada), NDGF (Denmark, Norway, Sweden), CC-IN2P3 (France), KIT/GridKA (Germany), INFN-CNAF (Italy), NL-T1 (Netherlands), PIC (Spain), ASGC (Taiwan), RAL (UK) and BNL (USA), the Tier-2 facilities worldwide and large non-WLCG resource providers. Major contributors of computing resources are listed in Ref.~\cite{ATL-SOFT-PUB-2021-003}.
 

\printbibliography

@article{LUCID2,
  author={G. Avoni and others},
  title={The new LUCID-2 detector for luminosity measurement and monitoring in ATLAS},
  journal={JINST},
  volume={13},
  number={07},
  pages={P07017},
  doi="10.1088/1748-0221/13/07/P07017",
  year={2018},
}

@article{Cowan:2010js,
    author = "Cowan, Glen and Cranmer, Kyle and Gross, Eilam and Vitells, Ofer",
    title = "{Asymptotic formulae for likelihood-based tests of new physics}",
    eprint = "1007.1727",
    archivePrefix = "arXiv",
    primaryClass = "physics.data-an",
    doi = "10.1140/epjc/s10052-011-1554-0",
    journal = "Eur. Phys. J. C",
    volume = "71",
    pages = "1554",
    year = "2011",
    note = "[Erratum: Eur.Phys.J.C 73, 2501 (2013)]"
}

@article{BELANGER1992201,
    title = {Probing quartic couplings of weak bosons through three vector production at a 500 GeV NLC},
    journal = {Phys. Lett. B},
    volume = {288},
    number = {1},
    pages = {201-209},
    year = {1992},
    %issn = {0370-2693},
    doi = {https://doi.org/10.1016/0370-2693(92)91978-I},
    %url = {https://www.sciencedirect.com/science/article/pii/037026939291978I},
    author = {G. Bélanger and F. Boudjema},
}

@article{Godfrey:1995,
    author = "Godfrey, Stephen",
    editor = "Baur, U. and Errede, S. and Mueller, T.",
    title = "{Quartic gauge boson couplings}",
    eprint = "hep-ph/9505252",
    archivePrefix = "arXiv",
    reportNumber = "OCIP-C-95-5",
    doi = "10.1063/1.49305",
    journal = "AIP Conf. Proc.",
    volume = "350",
    pages = "209--223",
    year = "1995"
}

@Article{Evans:2008zzb,
      author         = "Evans, Lyndon and Bryant, Philip",
      title          = "{LHC Machine}",
      journal        = "JINST",
      volume         = "3",
      pages          = "S08001",
      doi            = "10.1088/1748-0221/3/08/S08001",
      year           = "2008",
      SLACcitation   = "%%CITATION = JINST,3,S08001;%%",
}

@Article{Butterworth:2015oua,
      author         = "Butterworth, Jon and others",
      title          = "{PDF4LHC recommendations for LHC Run II}",
      journal        = "J. Phys. G",
      volume         = "43",
      year           = "2016",
      pages          = "023001",
      doi            = "10.1088/0954-3899/43/2/023001",
      eprint         = "1510.03865",
      archivePrefix  = "arXiv",
      primaryClass   = "hep-ph",
      reportNumber   = "OUTP-15-17P, SMU-HEP-15-12, TIF-UNIMI-2015-14,
                        LCTS-2015-27, CERN-PH-TH-2015-249",
      SLACcitation   = "%%CITATION = ARXIV:1510.03865;%%"
}

@article{Ball:2014uwa,
      author         = "Ball, Richard D. and others",
      title          = "{Parton distributions for the LHC run II}",
      collaboration  = "NNPDF",
      journal        = "JHEP",
      volume         = "04",
      year           = "2015",
      pages          = "040",
      doi            = "10.1007/JHEP04(2015)040",
      eprint         = "1410.8849",
      archivePrefix  = "arXiv",
      primaryClass   = "hep-ph",
      reportNumber   = "EDINBURGH-2014-15, IFUM-1034-FT, CERN-PH-TH-2013-253,
                        OUTP-14-11P, CAVENDISH-HEP-14-11",
      SLACcitation   = "%%CITATION = ARXIV:1410.8849;%%"
}

@article{Mrenna:2016sih,
     author         = "Mrenna, S. and Skands, P.",
     title          = "{Automated parton-shower variations in PYTHIA 8}",
     journal        = "Phys. Rev. D",
     volume         = "94",
     year           = "2016",
     pages          = "074005",
     doi            = "10.1103/PhysRevD.94.074005",
     eprint         = "1605.08352",
     archivePrefix  = "arXiv",
     primaryClass   = "hep-ph",
}

@Article{Sjostrand:2007gs,
	author    = "Sj{\"o}strand, T. and Mrenna, S. and Skands, P.",
	title     = "{A brief introduction to PYTHIA 8.1}",
	journal   = "Comput. Phys. Commun.",
	volume    = "178",
	year      = "2008",
	pages     = "852-867",
	eprint    = "0710.3820",
	archivePrefix = "arXiv",
	primaryClass  =  "hep-ph",
	doi       = "10.1016/j.cpc.2008.01.036",
	SLACcitation  = "%%CITATION = 0710.3820;%%"
}

@Article{Alwall:2014hca,
      author         = "Alwall, J. and Frederix, R. and Frixione, S. and Hirschi,
                        V. and Maltoni, F. and Mattelaer, O. and Shao, H. -S. and
                        Stelzer, T. and Torrielli, P. and Zaro, M.",
      title          = "{The automated computation of tree-level and
                        next-to-leading order differential cross sections, and
                        their matching to parton shower simulations}",
      journal        = "JHEP",
      volume         = "07",
      year           = "2014",
      pages          = "079",
      doi            = "10.1007/JHEP07(2014)079",
      eprint         = "1405.0301",
      archivePrefix  = "arXiv",
      primaryClass   = "hep-ph",
      reportNumber   = "CERN-PH-TH-2014-064, CP3-14-18, LPN14-066, MCNET-14-09,
                        ZU-TH-14-14",
      SLACcitation   = "%%CITATION = ARXIV:1405.0301;%%"
}

@Article{Gleisberg:2008ta,
      author         = "Gleisberg, T. and H{\"o}che, S. and Krauss, F. and Sch{\"o}nherr, M. and Schumann, S. and Siegert, F. and Winter J.",
      title          = "{Event generation with SHERPA 1.1}",
      journal        = "JHEP",
      volume         = "02",
      pages          = "007",
      doi            = "10.1088/1126-6708/2009/02/007",
      year           = "2009",
      eprint         = "0811.4622",
      archivePrefix  = "arXiv",
      primaryClass   = "hep-ph",
}

@Article{Hoeche:2009rj,
      author         = "H{\"o}che, Stefan and Krauss, Frank and Schumann, Steffen and Siegert, Frank",
      title          = "{QCD matrix elements and truncated showers}",
      journal        = "JHEP",
      volume         = "05",
      pages          = "053",
      doi            = "10.1088/1126-6708/2009/05/053",
      year           = "2009",
      eprint         = "0903.1219",
      archivePrefix  = "arXiv",
      primaryClass   = "hep-ph",
}

@article{Anastasiou:2016cez,
      author         = "Anastasiou, Charalampos and Duhr, Claude and Dulat, Falko
                        and Furlan, Elisabetta and Gehrmann, Thomas and Herzog,
                        Franz and Lazopoulos, Achilleas and Mistlberger, Bernhard",
      title          = "{High precision determination of the gluon fusion Higgs
                        boson cross-section at the LHC}",
      journal        = "JHEP",
      volume         = "05",
      year           = "2016",
      pages          = "058",
      doi            = "10.1007/JHEP05(2016)058",
      eprint         = "1602.00695",
      archivePrefix  = "arXiv",
      primaryClass   = "hep-ph",
      reportNumber   = "CP3-16-01, ZU-TH-27-15, NIKHEF-2016-004,
                        CERN-TH-2016-006",
      SLACcitation   = "%%CITATION = ARXIV:1602.00695;%%"
}

@article{Luisoni:2013cuh,
    author = "Luisoni, Gionata and Nason, Paolo and Oleari, Carlo and Tramontano, Francesco",
    title = "{\(HW^{\pm}/HZ\) + 0 and 1 jet at NLO with the POWHEG BOX interfaced to GoSam and their merging within MiNLO}",
    eprint = "1306.2542",
    archivePrefix = "arXiv",
    primaryClass = "hep-ph",
    doi = "10.1007/JHEP10(2013)083",
    journal = "JHEP",
    volume = "10",
    pages = "083",
    year = "2013"
}

@article{Agostinelli:2002hh,
      author         = "{GEANT4 Collaboration} and Agostinelli, S. and others",
      title          = "{\textsc{Geant4} -- a simulation toolkit}",
      journal        = "Nucl. Instrum. Meth. A",
      volume         = "506",
      year           = "2003",
      pages          = "250",
      doi            = "10.1016/S0168-9002(03)01368-8",
      reportNumber   = "SLAC-PUB-9350, FERMILAB-PUB-03-339",
      SLACcitation   = "%%CITATION = NUIMA,A506,250;%%"
}

@Article{PERF-2007-01,
    author         = "{ATLAS Collaboration}",
    title          = "{The ATLAS Experiment at the CERN Large Hadron Collider}",
    journal        = "JINST",
    volume         = "3",
    year           = "2008",
    pages          = "S08003",
    doi            = "10.1088/1748-0221/3/08/S08003",
    primaryClass   = "hep-ex",
}

@Article{SOFT-2010-01,
    author         = "{ATLAS Collaboration}",
    title          = "{The ATLAS Simulation Infrastructure}",
    journal        = "Eur. Phys. J. C",
    volume         = "70",
    year           = "2010",
    pages          = "823",
    doi            = "10.1140/epjc/s10052-010-1429-9",
    eprint         = "1005.4568",
    archivePrefix  = "arXiv",
    primaryClass   = "physics.ins-det",
}

@Article{STDM-2012-23,
    author         = "{ATLAS Collaboration}",
    title          = "{Measurement of the \(Z/\gamma^*\) boson transverse momentum distribution in \(pp\) collisions at \(\sqrt{s} = 7\,\text{TeV}\) with the ATLAS detector}",
    journal        = "JHEP",
    volume         = "09",
    year           = "2014",
    pages          = "145",
    doi            = "10.1007/JHEP09(2014)145",
    reportNumber   = "CERN-PH-EP-2014-075",
    eprint         = "1406.3660",
    archivePrefix  = "arXiv",
    primaryClass   = "hep-ex",
}

@Article{PERF-2014-03,
    author         = "{ATLAS Collaboration}",
    title          = "{Performance of pile-up mitigation techniques for jets in \(pp\) collisions at \(\sqrt{s} = 8\,\text{TeV}\) using the ATLAS detector}",
    journal        = "Eur. Phys. J. C",
    volume         = "76",
    year           = "2016",
    pages          = "581",
    doi            = "10.1140/epjc/s10052-016-4395-z",
    reportNumber   = "CERN-PH-EP-2015-206",
    eprint         = "1510.03823",
    archivePrefix  = "arXiv",
    primaryClass   = "hep-ex",
}

@Article{STDM-2014-05,
    author         = "{ATLAS Collaboration}",
    title          = "{Measurement of \(W^{\pm}W^{\pm}\) vector-boson scattering and limits on anomalous quartic gauge couplings with the ATLAS detector}",
    journal        = "Phys. Rev. D",
    volume         = "96",
    year           = "2017",
    pages          = "012007",
    doi            = "10.1103/PhysRevD.96.012007",
    reportNumber   = "CERN-EP-2016-167",
    eprint         = "1611.02428",
    archivePrefix  = "arXiv",
    primaryClass   = "hep-ex",
}

@Article{PERF-2015-09,
    author         = "{ATLAS Collaboration}",
    title          = "{Jet reconstruction and performance using particle flow with the ATLAS Detector}",
    journal        = "Eur. Phys. J. C",
    volume         = "77",
    year           = "2017",
    pages          = "466",
    doi            = "10.1140/epjc/s10052-017-5031-2",
    reportNumber   = "CERN-EP-2017-024",
    eprint         = "1703.10485",
    archivePrefix  = "arXiv",
    primaryClass   = "hep-ex",
}

@Article{STDM-2015-07,
    author         = "{ATLAS Collaboration}",
    title          = "{Search for triboson \(W^{\pm }W^{\pm }W^{\mp }\) production in \(pp\) collisions at \(\sqrt{s} = 8\,\text{TeV}\) with the ATLAS detector}",
    journal        = "Eur. Phys. J. C",
    volume         = "77",
    year           = "2017",
    pages          = "141",
    doi            = "10.1140/epjc/s10052-017-4692-1",
    reportNumber   = "CERN-EP-2016-172",
    eprint         = "1610.05088",
    archivePrefix  = "arXiv",
    primaryClass   = "hep-ex",
}

@Article{PERF-2016-06,
    author         = "{ATLAS Collaboration}",
    title          = "{Identification and rejection of pile-up jets at high pseudorapidity with the ATLAS detector}",
    journal        = "Eur. Phys. J. C",
    volume         = "77",
    year           = "2017",
    pages          = "580",
    doi            = "10.1140/epjc/s10052-017-5081-5",
    reportNumber   = "CERN-EP-2017-055",
    eprint         = "1705.02211",
    archivePrefix  = "arXiv",
    primaryClass   = "hep-ex",
    related        = "PERF-2016-06-err",
    relatedstring  = "Erratum:",
}

@Article{PERF-2016-07,
    author         = "{ATLAS Collaboration}",
    title          = "{Performance of missing transverse momentum reconstruction with the ATLAS detector using proton--proton collisions at \(\sqrt{s} = 13\,\text{TeV}\)}",
    journal        = "Eur. Phys. J. C",
    volume         = "78",
    year           = "2018",
    pages          = "903",
    doi            = "10.1140/epjc/s10052-018-6288-9",
    reportNumber   = "CERN-EP-2017-274",
    eprint         = "1802.08168",
    archivePrefix  = "arXiv",
    primaryClass   = "hep-ex",
}

@Article{TOPQ-2016-11,
    author         = "{ATLAS Collaboration}",
    title          = "{Measurement of the \(t\bar{t}Z\) and \(t\bar{t}W\) cross sections in proton--proton collisions at \(\sqrt{s} = 13\,\text{TeV}\) with the ATLAS detector}",
    journal        = "Phys. Rev. D",
    volume         = "99",
    year           = "2019",
    pages          = "072009",
    doi            = "10.1103/PhysRevD.99.072009",
    reportNumber   = "CERN-EP-2018-331",
    eprint         = "1901.03584",
    archivePrefix  = "arXiv",
    primaryClass   = "hep-ex",
}

@Article{TRIG-2016-01,
    author         = "{ATLAS Collaboration}",
    title          = "{Performance of the ATLAS trigger system in 2015}",
    journal        = "Eur. Phys. J. C",
    volume         = "77",
    year           = "2017",
    pages          = "317",
    doi            = "10.1140/epjc/s10052-017-4852-3",
    reportNumber   = "CERN-EP-2016-241",
    eprint         = "1611.09661",
    archivePrefix  = "arXiv",
    primaryClass   = "hep-ex",
}

@Article{HIGG-2017-02,
    author         = "{ATLAS Collaboration}",
    title          = "{Evidence for the associated production of the Higgs boson and a top quark pair with the ATLAS detector}",
    journal        = "Phys. Rev. D",
    volume         = "97",
    year           = "2018",
    pages          = "072003",
    doi            = "10.1103/PhysRevD.97.072003",
    reportNumber   = "CERN-EP-2017-281",
    eprint         = "1712.08891",
    archivePrefix  = "arXiv",
    primaryClass   = "hep-ex",
}

@Article{PERF-2017-01,
    author         = "{ATLAS Collaboration}",
    title          = "{Electron reconstruction and identification in the ATLAS experiment using the 2015 and 2016 LHC proton--proton collision data at \(\sqrt{s} = 13\,\text{TeV}\)}",
    journal        = "Eur. Phys. J. C",
    volume         = "79",
    year           = "2019",
    pages          = "639",
    doi            = "10.1140/epjc/s10052-019-7140-6",
    reportNumber   = "CERN-EP-2018-273",
    eprint         = "1902.04655",
    archivePrefix  = "arXiv",
    primaryClass   = "hep-ex",
}

@Article{PERF-2017-03,
    author         = "{ATLAS Collaboration}",
    title          = "{Electron and photon energy calibration with the ATLAS detector using 2015--2016 LHC proton--proton collision data}",
    journal        = "JINST",
    volume         = "14",
    year           = "2019",
    pages          = "P03017",
    doi            = "10.1088/1748-0221/14/03/P03017",
    reportNumber   = "CERN-EP-2018-296",
    eprint         = "1812.03848",
    archivePrefix  = "arXiv",
    primaryClass   = "hep-ex",
}

@Article{STDM-2017-06,
    author         = "{ATLAS Collaboration}",
    title          = "{Observation of Electroweak Production of a Same-Sign \(W\) Boson Pair in Association with Two Jets in \(pp\) Collisions at \(\sqrt{s} = 13\,\text{TeV}\) with the ATLAS Detector}",
    journal        = "Phys. Rev. Lett.",
    volume         = "123",
    year           = "2019",
    pages          = "161801",
    doi            = "10.1103/PhysRevLett.123.161801",
    reportNumber   = "CERN-EP-2019-008",
    eprint         = "1906.03203",
    archivePrefix  = "arXiv",
    primaryClass   = "hep-ex",
}

@Article{STDM-2017-22,
    author         = "{ATLAS Collaboration}",
    title          = "{Evidence for the production of three massive vector bosons with the ATLAS detector}",
    journal        = "Phys. Lett. B",
    volume         = "798",
    year           = "2019",
    pages          = "134913",
    doi            = "10.1016/j.physletb.2019.134913",
    reportNumber   = "CERN-EP-2019-041",
    eprint         = "1903.10415",
    archivePrefix  = "arXiv",
    primaryClass   = "hep-ex",
}

@Article{EGAM-2018-01,
    author         = "{ATLAS Collaboration}",
    title          = "{Electron and photon performance measurements with the ATLAS detector using the 2015--2017 LHC proton--proton collision data}",
    journal        = "JINST",
    volume         = "14",
    year           = "2019",
    pages          = "P12006",
    doi            = "10.1088/1748-0221/14/12/P12006",
    reportNumber   = "CERN-EP-2019-145",
    eprint         = "1908.00005",
    archivePrefix  = "arXiv",
    primaryClass   = "hep-ex",
}

@Article{FTAG-2018-01,
    author         = "{ATLAS Collaboration}",
    title          = "{ATLAS \(b\)-jet identification performance and efficiency measurement with \(t\bar{t}\) events in \(pp\) collisions at \(\sqrt{s} = 13\,\text{TeV}\)}",
    journal        = "Eur. Phys. J. C",
    volume         = "79",
    year           = "2019",
    pages          = "970",
    doi            = "10.1140/epjc/s10052-019-7450-8",
    reportNumber   = "CERN-EP-2019-132",
    eprint         = "1907.05120",
    archivePrefix  = "arXiv",
    primaryClass   = "hep-ex",
}

@Article{JETM-2018-05,
    author         = "{ATLAS Collaboration}",
    title          = "{Jet energy scale and resolution measured in proton--proton collisions at \(\sqrt{s} = 13\,\text{TeV}\) with the ATLAS detector}",
    journal        = "Eur. Phys. J. C",
    volume         = "81",
    year           = "2020",
    pages          = "689",
    doi            = "10.1140/epjc/s10052-021-09402-3",
    reportNumber   = "CERN-EP-2020-083",
    eprint         = "2007.02645",
    archivePrefix  = "arXiv",
    primaryClass   = "hep-ex",
}

@Article{MUON-2018-03,
    author         = "{ATLAS Collaboration}",
    title          = "{Muon reconstruction and identification efficiency in ATLAS using the full Run~2 \(pp\) collision data set at \(\sqrt{s} = 13\,\text{TeV}\)}",
    journal        = "Eur. Phys. J. C",
    volume         = "81",
    year           = "2021",
    pages          = "578",
    doi            = "10.1140/epjc/s10052-021-09233-2",
    reportNumber   = "CERN-EP-2020-199",
    eprint         = "2012.00578",
    archivePrefix  = "arXiv",
    primaryClass   = "hep-ex",
}

@Article{TOPQ-2018-01,
    author         = "{ATLAS Collaboration}",
    title          = "{Observation of the associated production of a top quark and a \(Z\) boson in \(pp\) collisions at \(\sqrt{s} = 13\,\text{TeV}\) with the ATLAS detector}",
    journal        = "JHEP",
    volume         = "07",
    year           = "2020",
    pages          = "124",
    doi            = "10.1007/JHEP07(2020)124",
    reportNumber   = "CERN-EP-2019-273",
    eprint         = "2002.07546",
    archivePrefix  = "arXiv",
    primaryClass   = "hep-ex",
}

@Article{TRIG-2018-05,
    author         = "{ATLAS Collaboration}",
    title          = "{Performance of electron and photon triggers in ATLAS during LHC Run~2}",
    journal        = "Eur. Phys. J. C",
    volume         = "80",
    year           = "2020",
    pages          = "47",
    doi            = "10.1140/epjc/s10052-019-7500-2",
    reportNumber   = "CERN-EP-2019-169",
    eprint         = "1909.00761",
    archivePrefix  = "arXiv",
    primaryClass   = "hep-ex",
}

@Booklet{ATL-SOFT-PUB-2021-001,
    author         = "{ATLAS Collaboration}",
    title          = "{The ATLAS Collaboration Software and Firmware}",
    howpublished   = "{ATL-SOFT-PUB-2021-001}",
    url            = "https://cds.cern.ch/record/2767187",
    year           = "2021",
}

@Booklet{ATL-SOFT-PUB-2021-003,
    author         = "{ATLAS Collaboration}",
    title          = "{ATLAS Computing Acknowledgements}",
    howpublished   = "{ATL-SOFT-PUB-2021-003}",
    url            = "https://cds.cern.ch/record/2776662",
}

@Article{CMS-SMP-17-013,
    author         = "{CMS Collaboration}",
    title          = "{Search for the production of \(W^\pm\)W\(^\pm\)W\(^\mp\) events at \(\sqrt{s} = 13\,\text{TeV}\)}",
    journal        = "Phys. Rev. D",
    volume         = "100",
    year           = "2019",
    pages          = "012004",
    doi            = "10.1103/PhysRevD.100.012004",
    reportNumber   = "CERN-EP-2019-074",
    eprint         = "1905.04246",
    archivePrefix  = "arXiv",
    primaryClass   = "hep-ex",
}

@Article{CMS-SMP-19-014,
    author         = "{CMS Collaboration}",
    title          = "{Observation of the production of three massive gauge bosons at \(\sqrt{s} = 13\,\text{TeV}\)}",
    journal        = "Phys. Rev. Lett.",
    volume         = "125",
    year           = "2020",
    pages          = "151802",
    doi            = "10.1103/PhysRevLett.125.151802",
    reportNumber   = "CERN-EP-2020-076",
    eprint         = "2006.11191",
    archivePrefix  = "arXiv",
    primaryClass   = "hep-ex",
}

@Booklet{ATLAS-CONF-2014-018,
    author         = "{ATLAS Collaboration}",
    title          = "{Tagging and suppression of pileup jets with the ATLAS detector}",
    howpublished   = "{ATLAS-CONF-2014-018}",
    url            = "https://cds.cern.ch/record/1700870",
    year           = "2014",
}

@Booklet{ATLAS-CONF-2018-006,
    author         = "{ATLAS Collaboration}",
    title          = "{Calibration of light-flavour \(b\)-jet mistagging rates using ATLAS proton--proton collision data at \(\sqrt{s} = 13~\text{TeV}\)}",
    howpublished   = "{ATLAS-CONF-2018-006}",
    url            = "https://cds.cern.ch/record/2314418",
    year           = "2018",
}

@Booklet{ATLAS-CONF-2018-038,
    author         = "{ATLAS Collaboration}",
    title          = "{Object-based missing transverse momentum significance in the ATLAS Detector}",
    howpublished   = "{ATLAS-CONF-2018-038}",
    url            = "https://cds.cern.ch/record/2630948",
    year           = "2018",
}

@Booklet{ATLAS-CONF-2019-021,
    author         = "{ATLAS Collaboration}",
    title          = "{Luminosity determination in \(pp\) collisions at \(\sqrt{s} = 13\,\text{TeV}\) using the ATLAS detector at the LHC}",
    howpublished   = "{ATLAS-CONF-2019-021}",
    url            = "https://cds.cern.ch/record/2677054",
    year           = "2019",
}

@Booklet{ATL-PHYS-PUB-2016-002,
    author         = "{ATLAS Collaboration}",
    title          = "{Multi-boson simulation for \(13~\text{TeV}\) ATLAS analyses}",
    howpublished   = "{ATL-PHYS-PUB-2016-002}",
    url            = "https://cds.cern.ch/record/2119986",
    year           = "2016",
}

@Booklet{ATL-DAQ-PUB-2019-001,
    author         = "{ATLAS Collaboration}",
    title          = "{Trigger Menu in 2018}",
    howpublished   = "{ATL-DAQ-PUB-2019-001}",
    url            = "https://cds.cern.ch/record/2693402",
    year           = "2019",
}

@article{LHCHiggsCrossSectionWorkingGroup:2016ypw,
    author = "de Florian, D. and others",
    collaboration = "LHC Higgs Cross Section Working Group",
    title = "{Handbook of LHC Higgs Cross Sections: 4. Deciphering the Nature of the Higgs Sector}",
    eprint = "1610.07922",
    archivePrefix = "arXiv",
    primaryClass = "hep-ph",
    reportNumber = "CERN-2017-002-M, CERN-2017-002",
    doi = "10.23731/CYRM-2017-002",
    volume = "2/2017",
    month = "10",
    year = "2016"
}

@article{Ren:2014bza,
    author = "Ren, Hong-Yu and Xia, Ling-Hao and Kuang, Yu-Ping",
    title = "{Model-independent Probe of anomalous heavy neutral Higgs bosons at the LHC}",
    eprint = "1404.6367",
    archivePrefix = "arXiv",
    primaryClass = "hep-ph",
    reportNumber = "TUHEP-TH-14180, TUHEP-TH-14180\&1",
    doi = "10.1103/PhysRevD.90.115002",
    journal = "Phys. Rev. D",
    volume = "90",
    number = "11",
    pages = "115002",
    year = "2014"
}

@article{Kuang:2015gka,
    author = "Kuang, Yu-Ping and Ren, Hong-Yu and Xia, Ling-Hao",
    title = "{Further investigation of the model-independent probe of heavy neutral Higgs bosons at LHC Run 2}",
    eprint = "1506.08007",
    archivePrefix = "arXiv",
    primaryClass = "hep-ph",
    reportNumber = "TUHEP-TH-151839",
    doi = "10.1088/1674-1137/40/2/023101",
    journal = "Chin. Phys. C",
    volume = "40",
    number = "2",
    pages = "023101",
    year = "2016"
}

@article{Aguilar-Saavedra:2017iso,
    author = "Aguilar-Saavedra, J. A.",
    title = "{Profile of multiboson signals}",
    eprint = "1703.06153",
    archivePrefix = "arXiv",
    primaryClass = "hep-ph",
    doi = "10.1007/JHEP05(2017)066",
    journal = "JHEP",
    volume = "05",
    pages = "066",
    year = "2017"
}

@article{Agashe:2017wss,
    author = "Agashe, Kaustubh and Collins, Jack H. and Du, Peizhi and Hong, Sungwoo and Kim, Doojin and Mishra, Rashmish K.",
    title = "{Dedicated Strategies for Triboson Signals from Cascade Decays of Vector Resonances}",
    eprint = "1711.09920",
    archivePrefix = "arXiv",
    primaryClass = "hep-ph",
    reportNumber = "CERN-TH-2017-233",
    doi = "10.1103/PhysRevD.99.075016",
    journal = "Phys. Rev. D",
    volume = "99",
    number = "7",
    pages = "075016",
    year = "2019"
}

@article{Agashe:2016kfr,
    author = "Agashe, Kaustubh S. and Collins, Jack and Du, Peizhi and Hong, Sungwoo and Kim, Doojin and Mishra, Rashmish K.",
    title = "{LHC Signals from Cascade Decays of Warped Vector Resonances}",
    eprint = "1612.00047",
    archivePrefix = "arXiv",
    primaryClass = "hep-ph",
    reportNumber = "CERN-TH-2016-247",
    doi = "10.1007/JHEP05(2017)078",
    journal = "JHEP",
    volume = "05",
    pages = "078",
    year = "2017"
}

@article{Agashe:2016rle,
    author = "Agashe, Kaustubh and Du, Peizhi and Hong, Sungwoo and Sundrum, Raman",
    title = "{Flavor Universal Resonances and Warped Gravity}",
    eprint = "1608.00526",
    archivePrefix = "arXiv",
    primaryClass = "hep-ph",
    doi = "10.1007/JHEP01(2017)016",
    journal = "JHEP",
    volume = "01",
    pages = "016",
    year = "2017"
}

@article{Aaboud:2018tkc,
    author = "{ATLAS Collaboration}",
    collaboration = "ATLAS",
    title = "{Performance of missing transverse momentum reconstruction with the ATLAS detector using proton-proton collisions at $\sqrt{s}$ = 13 TeV}",
    eprint = "1802.08168",
    archivePrefix = "arXiv",
    primaryClass = "hep-ex",
    reportNumber = "CERN-EP-2017-274",
    doi = "10.1140/epjc/s10052-018-6288-9",
    journal = "Eur. Phys. J. C",
    volume = "78",
    number = "11",
    pages = "903",
    year = "2018"
}

@article{Cranmer:2012sba,
    author = "Cranmer, Kyle and Lewis, George and Moneta, Lorenzo and Shibata, Akira and Verkerke, Wouter",
    collaboration = "ROOT",
    title = "{HistFactory: A tool for creating statistical models for use with RooFit and RooStats}",
    reportNumber = "CERN-OPEN-2012-016",
    journal = "CERN-OPEN-2012-016",
    month = "6",
    year = "2012"
}

@article{Dittmaier:2017bnh,
    author = "Dittmaier, Stefan and Huss, Alexander and Knippen, Gernot",
    title = "{Next-to-leading-order QCD and electroweak corrections to WWW production at proton-proton colliders}",
    eprint = "1705.03722",
    archivePrefix = "arXiv",
    primaryClass = "hep-ph",
    reportNumber = "FR-PHENO-2017-004",
    doi = "10.1007/JHEP09(2017)034",
    journal = "JHEP",
    volume = "09",
    pages = "034",
    year = "2017"
}

@inproceedings{Chen:2016:XST:2939672.2939785,
 author = {Chen, Tianqi and Guestrin, Carlos},
 title = {{XGBoost}: A Scalable Tree Boosting System},
 booktitle = {Proceedings of the 22nd ACM SIGKDD International Conference on Knowledge Discovery and Data Mining},
 series = {KDD '16},
 year = {2016},
 isbn = {978-1-4503-4232-2},
 location = {San Francisco, California, USA},
 pages = {785--794},
 numpages = {10},
 url = {http://doi.acm.org/10.1145/2939672.2939785},
 doi = {10.1145/2939672.2939785},
 acmid = {2939785},
 publisher = {ACM},
 address = {New York, NY, USA},
}
 
\clearpage
\FloatBarrier

\clearpage
 
\begin{flushleft}
\hypersetup{urlcolor=black}
{\Large The ATLAS Collaboration}

\bigskip

\AtlasOrcid[0000-0002-6665-4934]{G.~Aad}$^\textrm{\scriptsize 100}$,    
\AtlasOrcid[0000-0002-5888-2734]{B.~Abbott}$^\textrm{\scriptsize 126}$,    
\AtlasOrcid[0000-0002-7248-3203]{D.C.~Abbott}$^\textrm{\scriptsize 101}$,    
\AtlasOrcid[0000-0002-2788-3822]{A.~Abed~Abud}$^\textrm{\scriptsize 36}$,    
\AtlasOrcid[0000-0002-1002-1652]{K.~Abeling}$^\textrm{\scriptsize 53}$,    
\AtlasOrcid[0000-0002-2987-4006]{D.K.~Abhayasinghe}$^\textrm{\scriptsize 93}$,    
\AtlasOrcid[0000-0002-8496-9294]{S.H.~Abidi}$^\textrm{\scriptsize 29}$,    
\AtlasOrcid[0000-0002-9987-2292]{A.~Aboulhorma}$^\textrm{\scriptsize 35e}$,    
\AtlasOrcid[0000-0001-5329-6640]{H.~Abramowicz}$^\textrm{\scriptsize 158}$,    
\AtlasOrcid[0000-0002-1599-2896]{H.~Abreu}$^\textrm{\scriptsize 157}$
% ...
\bigskip
\\

% ...
$^{29}$Physics Department, Brookhaven National Laboratory, Upton NY; United States of America.\\
% ...
$^{35}$$^{(a)}$Facult\'e des Sciences Ain Chock, R\'eseau Universitaire de Physique des Hautes Energies - Universit\'e Hassan II, Casablanca;$^{(b)}$Facult\'{e} des Sciences, Universit\'{e} Ibn-Tofail, K\'{e}nitra;$^{(c)}$Facult\'e des Sciences Semlalia, Universit\'e Cadi Ayyad, LPHEA-Marrakech;$^{(d)}$LPMR, Facult\'e des Sciences, Universit\'e Mohamed Premier, Oujda;$^{(e)}$Facult\'e des sciences, Universit\'e Mohammed V, Rabat;$^{(f)}$Mohammed VI Polytechnic University, Ben Guerir; Morocco.\\
$^{36}$CERN, Geneva; Switzerland.\\
% ...
$^{53}$II. Physikalisches Institut, Georg-August-Universit\"{a}t G\"ottingen, G\"ottingen; Germany.\\
% ...
$^{93}$Department of Physics, Royal Holloway University of London, Egham; United Kingdom.\\
% ...
$^{100}$CPPM, Aix-Marseille Universit\'e, CNRS/IN2P3, Marseille; France.\\
$^{101}$Department of Physics, University of Massachusetts, Amherst MA; United States of America.\\
% ...
$^{126}$Homer L. Dodge Department of Physics and Astronomy, University of Oklahoma, Norman OK; United States of America.\\
% ...
$^{157}$Department of Physics, Technion, Israel Institute of Technology, Haifa; Israel.\\
$^{158}$Raymond and Beverly Sackler School of Physics and Astronomy, Tel Aviv University, Tel Aviv; Israel.\\
% ...
$^{a}$ Also at Borough of Manhattan Community College, City University of New York, New York NY; United States of America.\\
$^{b}$ Also at Bruno Kessler Foundation, Trento; Italy.\\
$^{c}$ Also at Center for High Energy Physics, Peking University; China.\\
$^{d}$ Also at Centro Studi e Ricerche Enrico Fermi; Italy.\\
$^{e}$ Also at CERN, Geneva; Switzerland.\\
$^{f}$ Also at D\'epartement de Physique Nucl\'eaire et Corpusculaire, Universit\'e de Gen\`eve, Gen\`eve; Switzerland.\\
% ...

\end{flushleft}


\end{document}